\documentclass[aps,prl,reprint,superscriptaddress,longbibliography]{revtex4-2}

\usepackage{amsmath,amssymb,bm}
\usepackage{graphicx}
\usepackage{hyperref}
\usepackage{xcolor}

\newcommand{\dd}{\mathrm{d}}

\newcommand{\Var}{\mathrm{Var}}
\newcommand{\KL}{D_{\mathrm{KL}}}

\begin{document}

\title{Entropic Reciprocity in Time-Reversed Young Interferometry}

\author{Jianming Wen}
\email{jwen7@binghamton.edu}
\affiliation{Department of Electrical and Computer Engineering, Binghamton University, Binghamton, New York 13902, USA}


\begin{abstract}
We show that time-reversed Young interferometry reorganizes, rather than reverses, optical entropy. A fixed detector conditions the reciprocal source--detector Green function and produces a source-label probability distribution. Marginal entropies in the standard and time-reversed geometries are generally unequal; the reciprocal invariant is instead the mutual information between source and detector coordinates. Near a destructive response, the conditioned source-label entropy can decrease while Fisher information for small phase, tilt, or defocus perturbations increases. The result identifies time-reversed Young interferometry as a source-space information processor with no analogue in ordinary detector-plane fringe readout.
\end{abstract}

\maketitle

Young's double-slit experiment is normally interpreted as the formation of a spatial intensity pattern at a detector~\cite{MandelWolf1995,BornWolf1999,GoodmanStatistical2015,GoodmanFourier2017}. A point source illuminates two paths, and the detector plane reveals the interference through a marginal distribution over the detector coordinate. In a time-reversed Young interferometer~\cite{Wen2025TRY,WenHybrid,Wen2026a,Wen2026b,Wen2026c,Wen2026d}, the operational roles are exchanged: a laterally addressable source plane replaces the usual observation screen, while the detector is held fixed. The resulting interference is not read as a detector-plane fringe pattern, but as a source-label response conditioned on a detection event. This geometry was recently introduced as a deterministic second-order interference effect in which the pattern plane coincides with the source plane rather than the detector plane \cite{Wen2025TRY,WenHybrid,Wen2026a,Wen2026b,Wen2026c,Wen2026d}.

This exchange raises a basic question: what, if anything, is reversed about the entropy and information flow? A literal reversal of thermodynamic entropy would be misleading. In an ideal reciprocal, lossless optical system, the field evolution is unitary, and the von Neumann entropy of a pure optical state is unchanged. The nontrivial effect is instead informational. Standard Young interference and time-reversed Young interference correspond to different reductions of the same source--detector propagation kernel. The former marginalizes over the source and reads out detector-space entropy; the latter fixes the detector and reads out conditioned source-label entropy. The central result of this Letter is that the reciprocal invariant is not a marginal entropy, but the mutual information between source and detector coordinates.

We describe the system by a scalar reciprocal propagation amplitude~\cite{BornWolf1999,Miller2000,Potton2004}
\begin{equation}
K_{\theta}(x,y),
\end{equation}
where $y$ is the source-plane coordinate, $x$ is the detector-plane coordinate, and $\theta$ denotes a small physical perturbation such as phase, tilt, defocus, wavelength shift, or geometric displacement. The experimentally relevant joint event distribution is
\begin{equation}
p_{\theta}(x,y)=\frac{\pi(y)\eta(x)\left|K_{\theta}(x,y)\right|^2}{\iint\dd x\dd y\,\pi(y)\eta(x) \left|K_{\theta}(x,y)\right|^2},\label{eq:joint}
\end{equation}
where $\pi(y)$ is the normalized source-label prior and $\eta(x)$ is the detector acceptance function. Equation~(2) is the common statistical object behind both geometries. It is useful to emphasize that $p_{\theta}(x,y)$ is the probability density for the accepted measurement record. The normalization denominator is therefore the total accepted detection weight for the chosen source prior and
detector acceptance. If one instead asks for information per launched photon, the no-detection outcome must be included as an additional event. This distinction becomes important near a dark response and is treated explicitly in the Supplemental Material~\cite{SM}. In the standard
Young readout, one observes the detector marginal
\begin{equation}
p_{\theta}^{\mathrm{std}}(x)=\int\dd y\,p_{\theta}(x,y).\label{eq:stdmarginal}
\end{equation}
In the time-reversed Young readout, one fixes a detector coordinate $x_0$ and observes the conditional source-label distribution
\begin{equation}
p_{\theta}^{\mathrm{TRY}}(y|x_0)=
\frac{\pi(y)\left|K_{\theta}(x_0,y)\right|^2}
{\int \dd y'\,\pi(y')\left|K_{\theta}(x_0,y')\right|^2}.\label{eq:tryconditional}
\end{equation}
Equation~(4) also gives a Bayesian interpretation of the TRY measurement. The source prior $\pi(y)$ is updated by the fixed-detector likelihood $|K_{\theta}(x_0,y)|^2$. Thus the TRY pattern is not merely
an intensity curve drawn in the source plane; it is the posterior source-label distribution selected by a detection event at $x_0$. This is the operational meaning of source-space conditioning in the present
work. The derivation of Eq.~(\ref{eq:joint}) from a full scalar diffraction model, together with the explicit reduction to Eqs.~(\ref{eq:stdmarginal}) and (\ref{eq:tryconditional}), is given in the Supplemental Material \cite{SM}.

Figure~\ref{fig:concept} summarizes the central structural point. Standard Young interference and time-reversed Young interference are not two unrelated experiments; they are two statistical reductions of the same reciprocal kernel $p_{\theta}(x,y)$. The standard geometry reads a detector marginal, whereas TRY reads a fixed-detector conditional slice. This distinction is the origin of the entropic asymmetry discussed below.

\begin{figure}[t]
\includegraphics[width=\columnwidth]{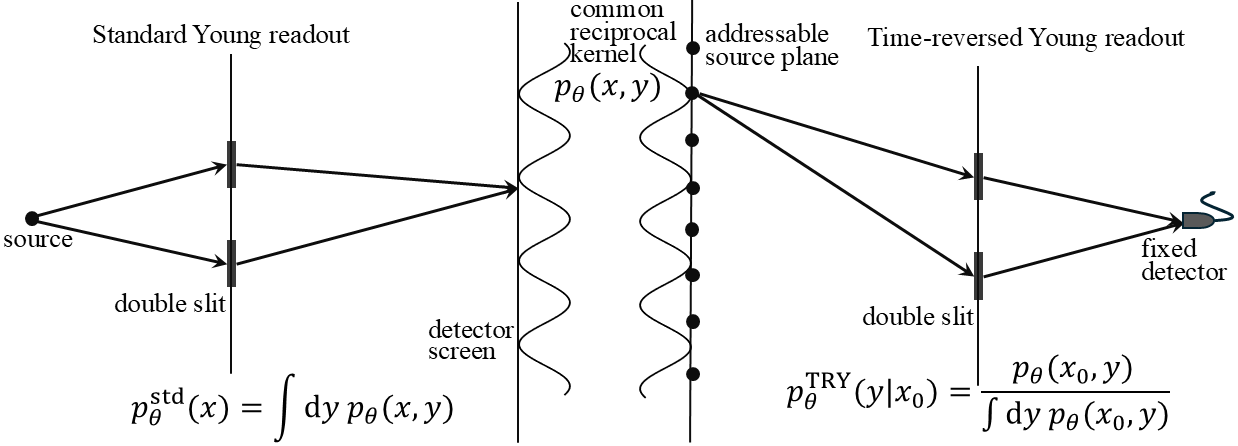}
\caption{
Conceptual comparison between standard Young readout and time-reversed Young readout. In the standard geometry, the detector screen measures the marginal distribution $p_{\theta}^{\mathrm{std}}(x)=\int\dd y\, p_{\theta}(x,y)$. In the time-reversed Young geometry, a fixed detector at $x_0$ selects the conditional source-label distribution $p_{\theta}^{\mathrm{TRY}}(y|x_0)=p_{\theta}(x_0,y)/\int\dd y\, p_{\theta}(x_0,y)$. Both readouts arise from the same reciprocal source--detector kernel, but they correspond to different statistical reductions: marginalization in detector space versus conditioning in source space.
}
\label{fig:concept}
\end{figure}

Equations~(\ref{eq:stdmarginal}) and (\ref{eq:tryconditional}) show why the two configurations are not entropy-equivalent. The standard experiment naturally measures a detector entropy,
\begin{equation}
H_X(\theta)=-\int\dd x\,p_{\theta}^{\mathrm{std}}(x)\ln p_{\theta}^{\mathrm{std}}(x),\label{eq:HX}
\end{equation}
whereas the time-reversed experiment measures a conditioned source entropy,
\begin{equation}
H_{Y|x_0}(\theta)=-\int\dd y\, p_{\theta}^{\mathrm{TRY}}(y|x_0)
\ln p_{\theta}^{\mathrm{TRY}}(y|x_0).\label{eq:HY}
\end{equation}
There is no general reciprocity principle requiring
$H_X=H_{Y|x_0}$. Indeed, they refer to different probability spaces: a detector marginal versus a source conditional. For continuous coordinates, $H_X$ and $H_{Y|x_0}$ are differential entropies and
therefore depend on the coordinate scale. In practice the source and detector coordinates are sampled by finite pixels, finite apertures, or
programmed source labels, so the measured quantities are discrete entropies of the recorded outcomes. The conclusions below do not rely on an absolute value of a differential entropy; they rely on relative
entropy, mutual information, and Fisher information, which have direct operational meanings for the measurement record.

The reciprocal quantity is the mutual information~\cite{Shannon1948,Kullback1951,Wehrl1978,CoverThomas2006}
\begin{equation}
I_{\theta}(X;Y)=\iint\dd x\dd y\,p_{\theta}(x,y)
\ln\frac{p_{\theta}(x,y)}{p_{\theta}(x)\,p_{\theta}(y)},\label{eq:MI}
\end{equation}
where $X$ and $Y$ denote the random variables associated with the detector coordinate $x$ and the source label $y$, respectively, while $p_{\theta}(x)$ and $p_{\theta}(y)$ are the corresponding marginals of $p_{\theta}(x,y)$. Equation~(\ref{eq:MI}) is symmetric under exchange of $X$ and $Y$. It measures the amount of source information contained in the detector coordinate, or equivalently the amount of detector information contained in the source label. Thus the time-reversed Young geometry does not invert the entropy itself; it reveals the complementary conditioning of the same joint optical information.

An equivalent form makes the role of conditioning especially transparent:
\begin{equation}
I_{\theta}(X;Y)=\int\dd x\,p_{\theta}(x)\,
D_{\mathrm{KL}}\!\left[p_{\theta}(y|x)\,\middle\|\,p_{\theta}(y)\right].
\label{eq:MI_as_average_KL}
\end{equation}
Thus the mutual information is the average information gained about the source label by observing the detector coordinate. Equation~\eqref{eq:KLidentity}
below is the corresponding single-outcome information gain for the fixed detector $x_0$, written relative to the programmed source prior in the uniform-prior case.

For a uniform source prior over an allowed source support of measure $A_y$, so that $\pi(y)=1/A_y$ on that support, Eq.~(\ref{eq:tryconditional}) gives a particularly transparent entropy identity,
\begin{equation}
\ln A_y-H_{Y|x_0}=\KL\!\left[
p_{\theta}^{\mathrm{TRY}}(y|x_0)\,\middle\|\,\pi(y)
\right].\label{eq:KLidentity}
\end{equation}
Here $A_y$ denotes the measure of the accessible source region, or, in a discretized source basis, the number of allowed source labels. The left-hand side of Eq.~\eqref{eq:KLidentity} is the entropy reduction of the source-label distribution caused by conditioning on a fixed detection event. The
right-hand side is the corresponding Kullback--Leibler information gain~\cite{Kullback1951,CoverThomas2006} relative to the programmed prior. Equation~\eqref{eq:KLidentity} is the first central result: TRY converts a single detector event into source-space information gain. Together with Eq.~(\ref{eq:MI_as_average_KL}), it shows that marginal entropy is not the object exchanged by reciprocity; rather, reciprocity acts on the joint source--detector information from which different conditional updates may be formed. A complete
derivation of Eq.~\eqref{eq:KLidentity} is given in the Supplemental Material~\cite{SM}.

Figure~\ref{fig:entropy} shows a representative numerical evaluation of Eqs.~(\ref{eq:HX})--(\ref{eq:MI}) for an illustrative reciprocal two-path model, with all model details deferred to the Supplemental Material \cite{SM}. The detector entropy $H_X$, conditioned source entropy $H_{Y|x_0}$, and mutual information $I_{\theta}(X;Y)$ evolve differently as a dimensionless reciprocal tuning parameter $\beta$ is varied. Their inequivalent behavior is the expected generic outcome: neither $H_X$ nor $H_{Y|x_0}$ is a reciprocal invariant. By contrast, Eq.~(\ref{eq:MI}) identifies the joint source--detector information as the proper reciprocal measure.

\begin{figure}[t]
\includegraphics[width=\columnwidth]{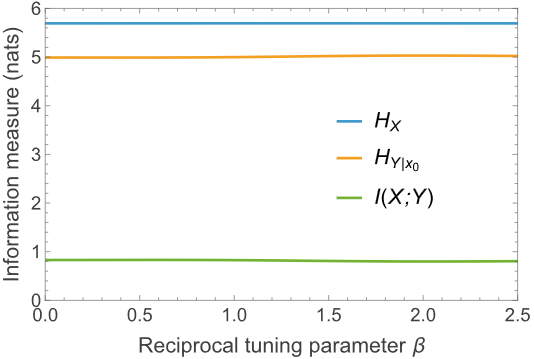}
\caption{
Illustrative numerical evaluation of Eqs.~\eqref{eq:HX}--\eqref{eq:MI} for a reciprocal two-path model as a dimensionless tuning parameter
$\beta$ is varied. The detector entropy $H_X$, conditioned source entropy $H_{Y|x_0}$, and mutual information $I(X;Y)$ are computed from the same joint kernel but from different statistical reductions. The important point is not equality of
their numerical scales, but their inequivalent status: $H_X$ and $H_{Y|x_0}$ are marginal or conditional entropies, whereas $I(X;Y)$ is the reciprocal source--detector information.
}
\label{fig:entropy}
\end{figure}

This entropy reduction is not merely descriptive. It is directly connected to local parameter sensitivity~\cite{Fisher1925,Rao1945,Cramer1946,Kay1993}. Define the fixed-detector response
\begin{equation}
R_{\theta}(y;x_0)=\left|K_{\theta}(x_0,y)\right|^2.
\label{eq:Rtheta}
\end{equation}
The score function for estimating $\theta$ from the TRY source-label distribution is
\begin{equation}
s_{\theta}^{\mathrm{TRY}}(y|x_0)=
\partial_{\theta}\ln p_{\theta}^{\mathrm{TRY}}(y|x_0)
=\partial_{\theta}\ln R_{\theta}(y;x_0)-\left\langle
\partial_{\theta}\ln R_{\theta}\right\rangle_{Y|x_0},
\label{eq:score}
\end{equation}
where $\langle\cdots\rangle_{Y|x_0}$ denotes averaging with respect to the conditional distribution $p_{\theta}^{\mathrm{TRY}}(y|x_0)$. The
second term in Eq.~\eqref{eq:score} is not an arbitrary offset. It is forced by normalization of the conditional distribution and ensures that
$\langle s_{\theta}^{\mathrm{TRY}}\rangle_{Y|x_0}=0$. Consequently, common-mode changes in the total detected response do not by themselves contribute to the conditional source-label Fisher information. What matters for $F_{\theta}^{\mathrm{TRY}}$ is the
reshaping of the source-label landscape caused by the perturbation. Therefore the Fisher information is
\begin{equation}
F_{\theta}^{\mathrm{TRY}}(x_0)=\Var_{Y|x_0}\!\left[
\partial_{\theta}\ln R_{\theta}(y;x_0)\right].
\label{eq:FI}
\end{equation}
Equation~(\ref{eq:FI}) is the second central result. The fixed detector does not simply collect less information than an image plane; rather, it defines
a reciprocal source-space statistical model whose sensitivity is governed by the variance of the source-space logarithmic response. Equation~\eqref{eq:FI} also gives a compact design rule. A perturbation is poorly estimated if $\partial_{\theta}\ln R_{\theta}(y;x_0)$ is nearly constant over the accepted source labels, because the variance then vanishes. Sensitivity is enhanced when the source basis samples regions where the
logarithmic response has large contrast. Source programming can therefore shape the score function directly: it is not simply a way to increase detected power, but a way to engineer the statistical
sensitivity of the measurement. The detailed derivation of
Eqs.~\eqref{eq:score} and \eqref{eq:FI}, together with the corresponding detector-plane
benchmark expressions, is provided in the Supplemental Material~\cite{SM}.

The most striking regime occurs near a destructive response~\cite{Bracewell1978,Lay2004}. Suppose that, near an operating point, the reciprocal response has the local form
\begin{equation}
R_{\theta}(y;x_0)\simeq\left|\epsilon(y)+\theta q(y)
\right|^2+b,\label{eq:nullmodel}
\end{equation}
where $\epsilon(y)$ is the residual nominal response, $q(y)$ is the first-order perturbation response, and $b$ is a small background floor. The enhancement implied by Eq.~(12) requires more than a small denominator. If $q(y)$ is locally proportional to $\epsilon(y)$ over the accepted source region, then $\partial_{\theta}\ln R_{\theta}$ is
approximately common-mode and the variance in Eq.~\eqref{eq:nullmodel} remains small. The useful regime is therefore a structured null: the nominal response is suppressed, while the perturbation response changes nonuniformly across the source labels. When $\epsilon(y)$ is small over the selected
source region and this structured condition is satisfied, the logarithmic derivative $\partial_{\theta}\ln R_{\theta}$ can become large and highly structured. The same conditioning that lowers the source-label entropy also increases the variance in Eq.~(\ref{eq:FI}), producing enhanced Fisher information per detected event. The enhancement is finite once the background floor, finite aperture, and total detection probability are included; these regularizations are essential and are treated in the Supplemental Material \cite{SM}. The important point is that TRY links dark-response conditioning, entropy reduction, and Fisher-information enhancement in a single reciprocal source-space readout.

\begin{figure}[t]
\includegraphics[width=\columnwidth]{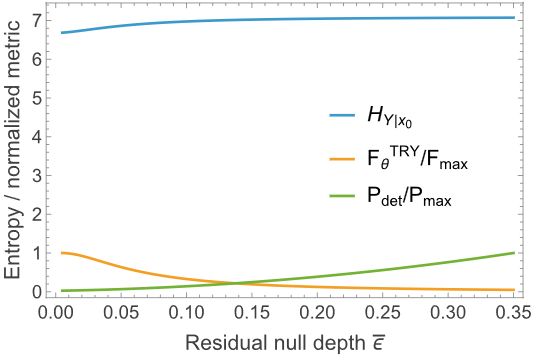}
\caption{
Illustrative null-response tradeoff based on Eq.~\eqref{eq:nullmodel}. As the residual null-depth control parameter $\bar{\epsilon}$ is reduced, the conditioned source entropy $H_{Y|x_0}$ decreases, while the TRY Fisher information per detected event increases. The normalized detection probability $P_{\mathrm{det}}/P_{\max}$ shows the complementary photon-budget cost of operating near a dark
response. Thus the null does not create unbounded information; it redistributes information into the conditional source-label record, with physical regularization by background, finite aperture, and normalization.
}
\label{fig:nullfi}
\end{figure}

Figure~\ref{fig:nullfi} illustrates this source-space concentration effect in a minimal regularized null model based on Eq.~(\ref{eq:nullmodel}). For compact display, we use $\bar{\epsilon}$ as a control parameter characterizing the residual null depth, for example an aperture-averaged amplitude mismatch. As $\bar{\epsilon}$ is reduced toward a deeper null, the conditioned source entropy decreases, while the TRY Fisher information per detected event increases. The normalized detection probability shown in the same figure makes clear that the gain in conditional sensitivity is accompanied by a cost in overall event rate. This is precisely the tradeoff that the Supplemental Material resolves quantitatively by distinguishing information per detected event from information per launched photon \cite{SM}.

This behavior has no direct analogue in ordinary detector-plane Young interference. A conventional detector measures the spatial consequence of interference after propagation. TRY instead uses the detector as a reciprocal filter that selects which source labels are compatible with a detection event. As a result, the source plane becomes an active measurement basis. A small phase, tilt, or defocus perturbation does not merely shift a fringe; it reshapes the conditional source-label distribution. In this sense, the time-reversed Young interferometer operates as a source-space information processor~\cite{Miller2000,Miller2012}.

The distinction can be summarized as follows. Standard Young interference maps path coherence into a detector marginal $p(x)$. Time-reversed Young interference maps the same reciprocal kernel into a
source conditional $p(y|x_0)$. Thermodynamic entropy is unchanged in the ideal lossless limit, but measurement entropy is redistributed by marginalization and conditioning. Marginal entropies are not reciprocal invariants. Mutual information is. Fisher information is not automatically invariant either, but it can be concentrated when the
source basis is chosen to make the perturbation score large and nonuniform. This is the sense in which TRY acts as a programmable source-space measurement architecture rather than a conventional
detector-plane fringe readout~\cite{Miller2000,Miller2012}. These results also clarify the role of thermal or partially coherent illumination. If the input field is mixed, its coherence entropy is determined by the eigenvalues of the corresponding coherence matrix~\cite{MandelWolf1995,GoodmanStatistical2015,Glauber1963Coherence,Wolf1982,Wolf2007,Refregier2006,Harling2024Locked,Lu2024CoherenceEntropy,Harling2024Isoentropy}. Under ideal unitary propagation, these eigenvalues are preserved. Entropy changes arise only through nonunitary operations: loss, aperture truncation, detector conditioning, coarse graining, or postselection. Thus the thermal language is meaningful only after specifying which degrees of freedom are traced out or conditioned upon. The Letter therefore focuses on the operational entropy of measured source--detector distributions, while the Supplemental Material gives the corresponding coherence-matrix treatment for partially coherent and thermal fields \cite{SM}.

In conclusion, time-reversed Young interferometry reveals a form of entropic reciprocity that is distinct from time reversal and distinct from thermodynamic irreversibility. The same reciprocal
Green function supports two inequivalent statistical reductions: detector-space marginalization in standard Young interference and source-space conditioning in TRY. The invariant connecting them is mutual information, while the operational advantage of TRY lies in the ability to sculpt conditioned source-label entropy and Fisher information using a fixed detector and a programmable
source basis. The resulting framework clarifies why a dark response can be useful without implying unbounded information: conditioning can enhance the Fisher information per detected event, while the full photon budget remains regularized by background, finite aperture, and detection probability. These results provide a compact theoretical basis for source-space null sensing, reciprocal autofocus, defocus estimation, wavelength
discrimination, and programmable fixed-detector interferometric metrology.

\begin{acknowledgments}
This work was partially supported by Binghamton University through Startup funds and Watson College through the internal award (No. 1201479).
\end{acknowledgments}

\end{document}


\title{Supplemental Material for ``Entropic Reciprocity in Time-Reversed Young Interferometry''}

\author{Jianming Wen}
\email{jwen7@binghamton.edu}
\affiliation{Department of Electrical and Computer Engineering, Binghamton University, Binghamton, New York 13902, USA}

\maketitle

This Supplemental Material provides the technical support for the main text. Sec.~\ref{secS:scope} states the modeling assumptions and connects the present formulation to the cited literature. Sec.~\ref{secS:kernel} derives the reciprocal source--detector kernel from scalar diffraction theory. Sec.~\ref{secS:corr} connects the source-label response to a labeled intensity-correlation language. Sec.~\ref{secS:entropy} derives the entropy, mutual-information, and Kullback--Leibler identities used in the main text. Sec.~\ref{secS:fisher} derives the time-reversed Young (TRY) Fisher-information formula and its detector-plane benchmark. Sec.~\ref{secS:null} treats the regularized null-response model. Sec.~\ref{secS:coherence} gives the coherence-matrix formulation for partially coherent and thermal fields. Sec.~\ref{secS:numerics} gives the numerical recipes used for the main-text and supplemental figures.

\section{Scope, assumptions, and connection to the cited literature}
\label{secS:scope}

The main text uses standard scalar diffraction and statistical optics, following the usual propagation and coherence notation of Goodman and Wolf~\cite{Goodman2015,Wolf2007}. The physical setting is a reciprocal linear optical system, so the source--detector Green function obeys the usual optical reciprocity constraints in the absence of nonreciprocal media, magneto-optic bias, or polarization-dependent asymmetry~\cite{Potton2004}. The starting experimental motivation is the time-reversed Young geometry originally introduced in Ref.~\cite{Wen2025TRY}, where a fixed detector and an addressable source plane replace the usual detector screen and fixed source.

The entropy and information quantities used in the main text are classical Shannon information measures~\cite{Shannon1948,Kullback1951,CoverThomas2006}. The Fisher-information formulas are standard classical estimation quantities~\cite{Fisher1925,Rao1945,Cramer1946,Kay1993}, here applied to the conditional source-label distribution selected by a fixed detector. The optical-correlation notation used below is consistent with the Glauber framework for field correlations~\cite{Glauber1963} and with the hybrid source-label correlation formulation of TRY~\cite{Wen2026Hybrid}. No two-detector Hanbury Brown--Twiss or ghost-imaging mechanism is assumed in the present derivation; the relevant statistical object is the single source--detector kernel and its marginal or conditional reductions. The novelty here is not the individual mathematical tools, but their coherent combination into a reciprocal source--detector entropy and Fisher-information theory for TRY.

Throughout this supplement, $x$ denotes a detector-plane coordinate, $y$ denotes a source-plane coordinate, and $\theta$ denotes a small physical perturbation such as phase, tilt, defocus, wavelength shift, or geometric displacement. The wave number is $k=2\pi/\lambda$, where $\lambda$ is the optical wavelength. We use natural logarithms, so entropies are measured in nats.

\section{Scalar diffraction model and reciprocal kernel}
\label{secS:kernel}

\subsection{Multi-slit propagation amplitude}
\label{subsecS:multislit}

Consider a scalar paraxial optical system with an addressable source plane, a slit plane, and a detector plane. Let $L_s$ be the distance from the source plane to the slit plane and $L_d$ the distance from the slit plane to the detector plane. Let $u$ denote the transverse coordinate in the slit plane. A general aperture transmission is written as $T(u)$. In the paraxial Fresnel approximation, propagation from $y$ to $u$ contributes a phase $\exp[ik(u-y)^2/(2L_s)]$, while propagation from $u$ to $x$ contributes $\exp[ik(x-u)^2/(2L_d)]$. Multiplying these two free-space kernels by the aperture transmission and integrating over the slit coordinate gives, up to an overall phase and normalization,
\begin{equation}
K_{\theta}(x,y)=\int\dd u\,T_{\theta}(u)
\exp\!\left[\frac{ik}{2L_s}(u-y)^2+\frac{ik}{2L_d}(x-u)^2\right],\label{eqS:generalK}
\end{equation}
where $T_{\theta}(u)$ may include a perturbation-dependent phase, amplitude, or aperture displacement. Equation~(\ref{eqS:generalK}) is the scalar Green-function amplitude (Eq.~(1)) used abstractly in the main text.

For an $N$-slit aperture with narrow slits centered at $u_m$, $m=1,\ldots,N$, one may write the aperture as a sum of localized transmissions. In the ideal narrow-slit limit this is equivalent to replacing $T_{\theta}(u)$ by a weighted sum of delta-like contributions. Equation~(\ref{eqS:generalK}) then reduces to
\begin{align}
K_{\theta}(x,y)=&C\sum_{m=1}^{N}a_m\exp\!\left[
\frac{ik}{2L_s}(u_m-y)^2+\frac{ik}{2L_d}(x-u_m)^2\right.\nonumber\\
&+i\phi_m(\theta)\bigg].\label{eqS:discreteK}
\end{align}
Here $C$ is a common normalization constant, $a_m$ is the complex amplitude transmission of the $m$th slit, and $\phi_m(\theta)$ is the perturbation-dependent phase at that slit. For a double slit,
\begin{equation}
u_{\pm}=\pm\,\frac{d}{2},
\label{eqS:doublepositions}
\end{equation}
where $d$ is the slit separation.

Equation~(\ref{eqS:discreteK}) is the explicit scalar-diffraction form underlying the abstract kernel $K_{\theta}(x,y)$ (Eq.~(1)) in the main text. It also shows why TRY is not merely a visual reversal of the ordinary Young pattern. The same amplitude contains both coordinates, but different experiments reduce it in different ways.

\subsection{Reciprocity}
\label{subsecS:reciprocity}

In a reciprocal scalar medium, the propagation amplitude from $y$ to $x$ is related to the amplitude obtained when source and detector roles are exchanged. In the present notation this means that the same Green-function kernel generates both the standard and TRY statistical models. If the geometry is physically exchanged, including the source--slit and slit--detector distances, the reversed amplitude has the same form as Eq.~(\ref{eqS:generalK}) with the roles of $x$ and $y$ interchanged. In general, this should be understood as a reciprocal Green-function statement, not as a literal equality $K(x,y)=K(y,x)$ in the same coordinate system unless the geometry, distances, polarizations, and normalizations are also symmetric.

This distinction is important for the main text. The word ``reciprocal'' refers to the source--detector optical kernel, while the statistical readout depends on whether one marginalizes over $y$ or conditions on a fixed $x_0$. Reciprocity therefore does not imply equality of the two measured entropies; it supplies the common kernel from which both statistical reductions are formed.

\subsection{Joint event distribution}
\label{subsecS:joint}

Let $\pi(y)$ be the normalized source-label prior,
\begin{equation}
\int\dd y\,\pi(y)=1,\label{eqS:priornorm}
\end{equation}
and let $\eta(x)\geq0$ be the detector acceptance function. A pointlike detector at $x_0$ corresponds to the limiting case $\eta(x)\rightarrow\delta(x-x_0)$, while a finite detector has a finite-width acceptance. The normalized joint distribution for an accepted measurement record with source label $y$ and detector coordinate $x$ is
\begin{equation}
p_{\theta}(x,y)=\frac{\pi(y)\eta(x)|K_{\theta}(x,y)|^2}{Z_{\theta}},\label{eqS:joint}
\end{equation}
where
\begin{equation}
Z_{\theta}=\int\dd x\dd y\,\pi(y)\eta(x) |K_{\theta}(x,y)|^2.\label{eqS:normalization}
\end{equation}
The constant $Z_{\theta}$ is the total accepted detection weight for the chosen source prior and detector aperture. Thus, Eq.~(\ref{eqS:joint}) describes the probability density conditioned on being in the accepted measurement record. If information is counted per launched photon, the no-detection outcome must be included separately; that photon-budget distinction is derived in Sec.~\ref{subsecS:launched}.

The standard Young readout is the detector marginal
\begin{equation}
p_{\theta}^{\mathrm{std}}(x)=\int\dd y\,p_{\theta}(x,y).\label{eqS:stdmarginal}
\end{equation}
The TRY readout at fixed detector coordinate $x_0$ is the conditional source-label distribution
\begin{equation}
p_{\theta}^{\mathrm{TRY}}(y|x_0)=
\frac{\pi(y)|K_{\theta}(x_0,y)|^2}
{\int\dd y'\,\pi(y')|K_{\theta}(x_0,y')|^2}.
\label{eqS:tryconditional}
\end{equation}
Equation~(\ref{eqS:tryconditional}) follows directly from Bayes normalization of the fixed-detector slice of Eq.~(\ref{eqS:joint}). For a finite detector window centered near $x_0$, the numerator $|K_{\theta}(x_0,y)|^2$ should be replaced by the acceptance-weighted response $\int\dd x\,\eta_{x_0}(x)|K_{\theta}(x,y)|^2$. Equations~(\ref{eqS:stdmarginal}) and (\ref{eqS:tryconditional}) are Eqs.~(3) and (4) of the main text.

\begin{figure}[hbpt]
\includegraphics[width=\columnwidth]{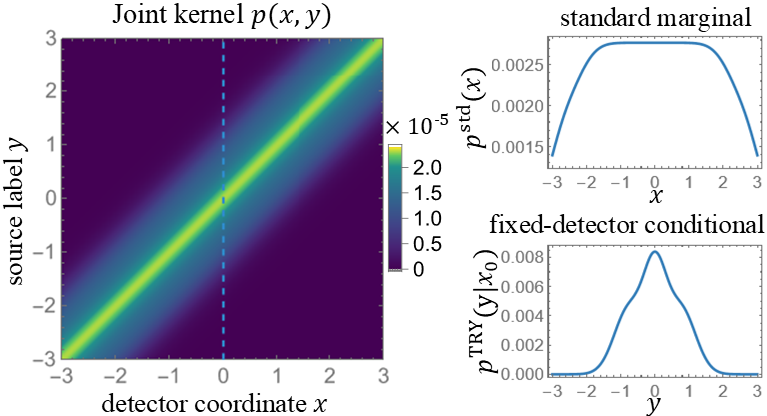}
\caption{
Joint-kernel view of the two statistical reductions. The left panel shows an illustrative normalized reciprocal kernel $p(x,y)$. The upper-right panel shows the standard detector marginal $p^{\mathrm{std}}(x)=\int \dd y\,p(x,y)$. The lower-right panel shows the fixed-detector conditional source distribution $p^{\mathrm{TRY}}(y|x_0)$ obtained from the vertical slice at $x_0=0$. The example is dimensionless and is used only to visualize the statistical structure of Eqs.~(\ref{eqS:stdmarginal}) and (\ref{eqS:tryconditional}).
}
\label{figS:kernel}
\end{figure}

Figure~\ref{figS:kernel} visualizes these two reductions. The left panel shows a representative joint kernel $p(x,y)$. The vertical dashed line marks the fixed detector $x_0=0$ used to form the TRY conditional distribution. The upper-right panel integrates the joint kernel over source labels to obtain the standard detector marginal. The lower-right panel takes a fixed-detector slice and normalizes it over source labels. Thus Fig.~\ref{figS:kernel} should be read as a map of the statistical operations in Eqs.~(\ref{eqS:stdmarginal}) and (\ref{eqS:tryconditional}), not as a claim about a unique experimental parameter set.

\subsection{Discrete source labels}
\label{subsecS:discrete}

Many TRY implementations are naturally discrete because the source plane is scanned, pixelated, or digitally programmed. If the allowed source labels are $y_j$, $j=1,\ldots,M$, with probabilities $\pi_j$, then Eq.~(\ref{eqS:tryconditional}) accordingly becomes
\begin{equation}
p_{\theta,j}^{\mathrm{TRY}}(x_0)=\frac{\pi_j |K_{\theta}(x_0,y_j)|^2}{\sum_{\ell=1}^{M} \pi_{\ell}|K_{\theta}(x_0,y_{\ell})|^2}.
\label{eqS:discreteTRY}
\end{equation}
All entropy identities below have direct discrete analogues. In fact, the discrete form is often the most experimentally robust because it avoids coordinate-unit ambiguities in differential entropy and corresponds directly to a finite measurement record.

\section{Connection to labeled intensity-correlation language}
\label{secS:corr}

The TRY readout can also be written in a correlation form. Let $\hat{E}^{(+)}(x_0)$ and $\hat{E}^{(-)}(x_0)$ denote the positive- and negative-frequency field operators at the fixed detector. Let $\hat{\Pi}_y$ denote the source-label projector, or, in a classical source-label implementation, the event label associated with the source coordinate $y$. A source-resolved intensity-label response has the schematic form~\cite{Wen2026Hybrid}
\begin{equation}
G^{(1,1)}(x_0;y)=\left\langle\hat{E}^{(-)}(x_0)\,
\hat{\Pi}_y\,\hat{E}^{(+)}(x_0)\right\rangle.
\label{eqS:hybridcorr}
\end{equation}
In a classical scanned-source realization, Eq.~(\ref{eqS:hybridcorr}) reduces operationally to the detected intensity at $x_0$ conditioned on the source label $y$,
\begin{equation}
G^{(1,1)}(x_0;y)\propto\pi(y)|K_{\theta}(x_0,y)|^2.\label{eqS:hybridclassical}
\end{equation}
After normalization over $y$, Eq.~(\ref{eqS:hybridclassical}) gives Eq.~(\ref{eqS:tryconditional}). This hybrid correlation notation is useful because it makes the source label part of the measurement record. It should not be interpreted as requiring a two-detector correlation experiment; for the present entropy theory, the essential object is the source-resolved fixed-detector response and its normalized conditional statistical distribution.

\section{Entropy identities and mutual information}
\label{secS:entropy}

\subsection{Detector entropy and source-label entropy}
\label{subsecS:entropies}

The standard detector entropy is
\begin{equation}
H_X(\theta)=-\int\dd x\,p_{\theta}^{\mathrm{std}}(x)\ln p_{\theta}^{\mathrm{std}}(x),\label{eqS:HX}
\end{equation}
where $X$ is the random variable associated with the detector coordinate. The fixed-detector source-label entropy is
\begin{equation}
H_{Y|x_0}(\theta)=-\int\dd y\,
p_{\theta}^{\mathrm{TRY}}(y|x_0)
\ln p_{\theta}^{\mathrm{TRY}}(y|x_0),
\label{eqS:HY}
\end{equation}
where $Y$ is the random variable associated with the source label. These are Eqs.~(5) and (6) of the main text.

For continuous coordinates, Eqs.~(\ref{eqS:HX}) and (\ref{eqS:HY}) are differential entropies and therefore depend on coordinate units. This does not affect the main conclusion because the reciprocal information statement is formulated using mutual information or Kullback--Leibler divergence~\cite{Kullback1951,CoverThomas2006}, both of which are invariant under smooth coordinate reparametrization when the probability densities transform properly. In a pixelated experiment, Eqs.~(\ref{eqS:HX}) and (\ref{eqS:HY}) are ordinary discrete Shannon entropies of the measured outcomes.

\subsection{Mutual information as reciprocal invariant}
\label{subsecS:MI}

The mutual information of the joint distribution $p_{\theta}(x,y)$ is
\begin{equation}
I_{\theta}(X;Y)=\int\dd x\dd y\,p_{\theta}(x,y)
\ln\frac{p_{\theta}(x,y)}{p_{\theta}(x)p_{\theta}(y)}.\label{eqS:MI}
\end{equation}
Using $p_{\theta}(x,y)=p_{\theta}(x)p_{\theta}(y|x)$, Eq.~(\ref{eqS:MI}) can be rewritten as
\begin{align}
I_{\theta}(X;Y)&=\int\dd x\,p_{\theta}(x)\int\dd y\,p_{\theta}(y|x)\ln\frac{p_{\theta}(y|x)}{p_{\theta}(y)}\nonumber\\
&=\int\dd x\,p_{\theta}(x)\KL\!\left[p_{\theta}(y|x)\middle\|p_{\theta}(y)\right].
\label{eqS:MIaverageKL}
\end{align}
Thus the mutual information is the average information gained about the source label by observing the detector coordinate. Equivalently,
\begin{equation}
I_{\theta}(X;Y)=H(X)-H(X|Y)=H(Y)-H(Y|X).
\label{eqS:MIidentities}
\end{equation}
Equation~(\ref{eqS:MIidentities}) makes the source--detector symmetry explicit. Marginal entropies such as $H(X)$ and conditioned single-slice entropies such as $H(Y|X=x_0)$ are not generally equal, but $I(X;Y)$ is symmetric under exchange of the random variables. This is the precise sense in which mutual information, rather than a marginal entropy, is the reciprocal information measure.

\subsection{Entropy reduction as KL information gain}
\label{subsecS:KL}

Assume first that the source prior is uniform over a support of measure $A_y$,
\begin{equation}
\pi(y)=\frac{1}{A_y}.\label{eqS:uniformprior}
\end{equation}
The Kullback--Leibler divergence from the conditional source-label distribution to the prior is
\begin{align}
&\KL\!\left[
p_{\theta}^{\mathrm{TRY}}(y|x_0)\middle\|\pi(y)
\right]
=\int\dd y\,p_{\theta}^{\mathrm{TRY}}(y|x_0)
\ln\frac{p_{\theta}^{\mathrm{TRY}}(y|x_0)}
{\pi(y)}\nonumber\\
&\quad=\int\dd y\,p_{\theta}^{\mathrm{TRY}}(y|x_0)
\ln p_{\theta}^{\mathrm{TRY}}(y|x_0)+\ln A_y
\nonumber\\
&\quad=\ln A_y-H_{Y|x_0}.\label{eqS:KLidentity}
\end{align}
In the second line we used $\pi(y)=1/A_y$ and $\int\dd y\,p_{\theta}^{\mathrm{TRY}}(y|x_0)=1$. This proves the entropy-reduction identity (Eq.~(9)) used in the main text. Equation~(\ref{eqS:KLidentity}) is the fixed-detector, single-outcome version of the average KL identity in Eq.~(\ref{eqS:MIaverageKL}); it quantifies how much the detection event at $x_0$ updates the source-label distribution relative to the programmed prior.

For a nonuniform prior, the corresponding identity is
\begin{equation}
\KL\!\left[p_{\theta}^{\mathrm{TRY}}(y|x_0)\middle\|\pi(y)\right]=-H_{Y|x_0}-\left\langle\ln\pi(y)\right\rangle_{Y|x_0}.
\label{eqS:nonuniformKL}
\end{equation}
Thus, the simple entropy-reduction form $(\ln A_y-H_{Y|x_0})$ is special to a uniform prior, but the interpretation as information gain relative to the prior remains general.

\begin{figure}[t]
\includegraphics[width=\columnwidth]{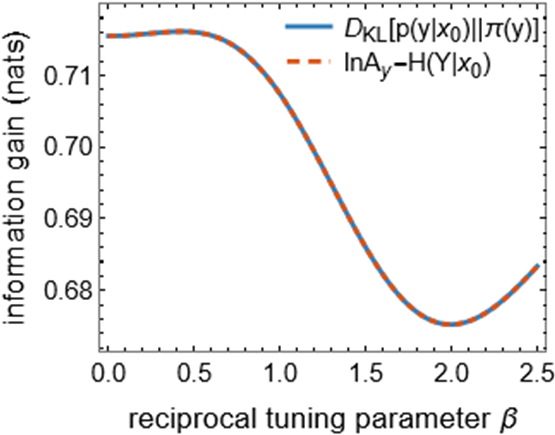}
\caption{
Numerical check of Eq.~(\ref{eqS:KLidentity}) for a uniform discrete source prior. The entropy reduction $\ln A_y-H_{Y|x_0}$ and the Kullback--Leibler divergence $\KL[p(y|x_0)\|\pi(y)]$ coincide for all values of the reciprocal tuning parameter $\beta$. The overlap of the two curves verifies that the fixed detector converts a detection event into source-space information gain.
}
\label{figS:KL}
\end{figure}

Figure~\ref{figS:KL} verifies Eq.~(\ref{eqS:KLidentity}) numerically for the same class of dimensionless two-path kernels used in the main text. The two plotted curves coincide because they are two representations of the same quantity. Their variation with the numerical tuning parameter $\beta$ is model dependent and should not be read as universal. The universal statement is the identity itself: the entropy reduction of a uniform source prior is exactly the KL information gain produced by fixed-detector conditioning.

\section{Fisher information in the TRY and standard readouts}
\label{secS:fisher}

\subsection{TRY score function}
\label{subsecS:score}

Define the fixed-detector response
\begin{equation}
R_{\theta}(y;x_0)=|K_{\theta}(x_0,y)|^2.
\label{eqS:R}
\end{equation}
The TRY conditional probability density can then be written as
\begin{equation}
p_{\theta}^{\mathrm{TRY}}(y|x_0)=
\frac{\pi(y)R_{\theta}(y;x_0)}{A_{\theta}(x_0)},
\label{eqS:ptrynorm}
\end{equation}
where
\begin{equation}
A_{\theta}(x_0)=\int\dd y\,\pi(y)R_{\theta}(y;x_0)
\label{eqS:Atheta}
\end{equation}
is the total fixed-detector response for the chosen source prior. If $\pi(y)$ is independent of $\theta$, taking the logarithm of Eq.~(\ref{eqS:ptrynorm}) gives
\begin{equation}
\ln p_{\theta}^{\mathrm{TRY}}(y|x_0)=
\ln\pi(y)+\ln R_{\theta}(y;x_0)-\ln A_{\theta}(x_0).\label{eqS:logptry}
\end{equation}
Differentiating Eq.~(\ref{eqS:logptry}) with respect to $\theta$ gives
\begin{equation}
\partial_{\theta}\ln p_{\theta}^{\mathrm{TRY}}(y|x_0)=\partial_{\theta}\ln R_{\theta}(y;x_0)
-\partial_{\theta}\ln A_{\theta}(x_0).
\label{eqS:score1}
\end{equation}
The second term is determined by normalization. Indeed,
\begin{align}
\partial_{\theta}\ln A_{\theta}(x_0)&=\frac{1}{A_{\theta}(x_0)}\int\dd y\,\pi(y) \partial_{\theta}R_{\theta}(y;x_0)\nonumber\\
&=\int\dd y\,\frac{\pi(y)R_{\theta}(y;x_0)}{A_{\theta}(x_0)}\frac{\partial_{\theta}R_{\theta}(y;x_0)}{R_{\theta}(y;x_0)}\nonumber\\
&=\int\dd y\,p_{\theta}^{\mathrm{TRY}}(y|x_0)
\partial_{\theta}\ln R_{\theta}(y;x_0)\nonumber\\
&=\left\langle\partial_{\theta}\ln R_{\theta}
\right\rangle_{Y|x_0}.\label{eqS:normderivative}
\end{align}
Combining Eqs.~(\ref{eqS:score1}) and (\ref{eqS:normderivative}) gives the TRY score
\begin{equation}
s_{\theta}^{\mathrm{TRY}}(y|x_0)=
\partial_{\theta}\ln R_{\theta}(y;x_0)-\left\langle\partial_{\theta}\ln R_{\theta}
\right\rangle_{Y|x_0},\label{eqS:score}
\end{equation}
which is Eq.~(11) in the main text. The subtraction in Eq.~(\ref{eqS:score}) is therefore not arbitrary. It is forced by the normalization of the conditional distribution and guarantees that $\langle s_{\theta}^{\mathrm{TRY}}\rangle_{Y|x_0}=0$. Physically, a common-mode change in the total detected response is removed from the conditional source-label statistics; what remains is the source-dependent reshaping of the normalized response.

\subsection{TRY Fisher information}
\label{subsecS:FITRY}

The Fisher information of the conditional source-label distribution is
\begin{align}
F_{\theta}^{\mathrm{TRY}}(x_0)&=\int\dd y\,
p_{\theta}^{\mathrm{TRY}}(y|x_0)\left[
s_{\theta}^{\mathrm{TRY}}(y|x_0)\right]^2
\nonumber\\
&=\Var_{Y|x_0}\!\left[\partial_{\theta}\ln R_{\theta}(y;x_0)\right].\label{eqS:FITRY}
\end{align}
This proves the Fisher-information expression (12) used in the main text. Equation~(\ref{eqS:FITRY}) shows that TRY sensitivity is controlled by the source-space variation of the logarithmic detector response. If $\partial_{\theta}\ln R_{\theta}(y;x_0)$ is nearly constant over the accepted source labels, the variance is small even if the total detected power changes. Sensitivity is enhanced when the source basis samples regions where the logarithmic response has large contrast. Thus source programming is not merely a way to increase detected power; it can engineer the score function of the measurement.

\subsection{Detector-plane benchmark}
\label{subsecS:stdFI}

The corresponding standard detector-plane response is
\begin{equation}
I_{\theta}^{\mathrm{std}}(x)=\int\dd y\,\pi(y) |K_{\theta}(x,y)|^2.\label{eqS:Istd}
\end{equation}
After normalization over the detector coordinate,
\begin{equation}
p_{\theta}^{\mathrm{std}}(x)=
\frac{\eta(x)I_{\theta}^{\mathrm{std}}(x)}
{\int\dd x'\,\eta(x')I_{\theta}^{\mathrm{std}}(x')}.\label{eqS:pstdnorm}
\end{equation}
If $\eta(x)$ is independent of $\theta$, the same normalization argument used above gives the standard detector-plane score
\begin{equation}
s_{\theta}^{\mathrm{std}}(x)=\partial_{\theta}\ln I_{\theta}^{\mathrm{std}}(x)-\left\langle
\partial_{\theta}\ln I_{\theta}^{\mathrm{std}}
\right\rangle_X,\label{eqS:scorestd}
\end{equation}
and the corresponding Fisher information is
\begin{equation}
F_{\theta}^{\mathrm{std}}=\Var_X\!\left[
\partial_{\theta}\ln I_{\theta}^{\mathrm{std}}(x)
\right].\label{eqS:FIstd}
\end{equation}
The TRY and standard expressions are structurally similar but are evaluated on different probability spaces. Equation~(\ref{eqS:FITRY}) is a variance over the source label conditioned on a fixed detector, while Eq.~(\ref{eqS:FIstd}) is a variance over detector coordinates. This difference is the Fisher-information analogue of the entropy distinction between $H_{Y|x_0}$ and $H_X$.

\subsection{Per detected event versus per launched photon}
\label{subsecS:launched}

The main text discusses Fisher information per detected event. For a complete photon budget, the detection probability must also be included. Let $r_{\theta}(y)$ be the probability for a launched photon with source label $y$ to be detected at the fixed detector, apart from the source prior. The total detection probability is
\begin{equation}
P_{\mathrm{det}}(\theta)=\int\dd y\,\pi(y) r_{\theta}(y).\label{eqS:Pdet}
\end{equation}
The conditional distribution of detected source labels is
\begin{equation}
q_{\theta}(y|\mathrm{det})=\frac{\pi(y)r_{\theta}(y)}{P_{\mathrm{det}}(\theta)}.\label{eqS:qdet}
\end{equation}
A launched photon produces either a no-detection event with probability $1-P_{\mathrm{det}}$ or a detected source-label record distributed according to $P_{\mathrm{det}}q_{\theta}(y|\mathrm{det})$. The Fisher information of this full launched-photon record is therefore
\begin{equation}
F_{\theta}^{\mathrm{launch}}=P_{\mathrm{det}}F_{\theta}^{\mathrm{cond}}+\frac{[\partial_{\theta}P_{\mathrm{det}}(\theta)]^2}{P_{\mathrm{det}}(\theta)[1-P_{\mathrm{det}}(\theta)]},\label{eqS:FIlaunch}
\end{equation}
where $F_{\theta}^{\mathrm{cond}}$ is the Fisher information of the conditional distribution $q_{\theta}(y|\mathrm{det})$. The first term is the information in the distribution of detected labels, weighted by the probability that a detection occurs. The second term is the count-rate information contained in whether the photon was detected at all. If no-detection events are not recorded or if the analysis is conditioned on successful detections, the relevant quantity is $F_{\theta}^{\mathrm{cond}}$. If one compares information per launched photon while ignoring the count-rate information, the conservative conditional contribution is $P_{\mathrm{det}}F_{\theta}^{\mathrm{cond}}$. This distinction is essential near a dark response, where conditional sensitivity can be high while the event rate is low.

\section{Regularized null-response model}
\label{secS:null}

\subsection{Local expansion}
\label{subsecS:nullexpansion}

Near a destructive response, the fixed-detector amplitude may be expanded as
\begin{equation}
\mathcal{A}_{\theta}(y;x_0)\simeq
\epsilon(y)+\theta q(y),\label{eqS:nullamplitude}
\end{equation}
where $\epsilon(y)$ is the residual nominal amplitude and $q(y)$ is the first-order perturbation response. We use the calligraphic symbol $\mathcal{A}_{\theta}$ for the local amplitude to avoid confusion with the normalization factor $A_{\theta}(x_0)$ in Eq.~(\ref{eqS:Atheta}). The measured response is modeled as
\begin{equation}
R_{\theta}(y;x_0)=|\epsilon(y)+\theta q(y)|^2+b,
\label{eqS:nullresponse}
\end{equation}
where $b>0$ is a background floor. This is the regularized null-response model (13) used in the main text.

For a real scalar perturbation $\theta$, differentiation of Eq.~(\ref{eqS:nullresponse}) gives, at $\theta=0$,
\begin{equation}
\partial_{\theta}\ln R_{\theta}(y;x_0) \big|_{\theta=0}=\frac{2\,\mathrm{Re}\!\left[\epsilon^*(y)q(y)\right]}{|\epsilon(y)|^2+b}.
\label{eqS:nulllogderivative}
\end{equation}
Equation~(\ref{eqS:nulllogderivative}) shows the regularized nature of the enhancement. If $|\epsilon|^2\gg b$, the logarithmic derivative scales roughly as $1/|\epsilon|$. If $|\epsilon|^2\ll b$, the background floor suppresses the logarithmic derivative. Therefore, a physical system does not produce unbounded Fisher information at an exact null. Instead, the strongest response occurs in the background-limited crossover region.

The null must also be structured. If $q(y)$ is locally proportional to $\epsilon(y)$ over the accepted source region, then Eq.~(\ref{eqS:nulllogderivative}) is approximately common-mode and the variance in Eq.~(\ref{eqS:FITRY}) remains small. The useful regime is therefore one in which the nominal response is suppressed while the perturbation response changes nonuniformly across the source labels. This is the source-space mechanism behind the entropy--Fisher concentration described in the main text.

\subsection{Finite Fisher information}
\label{subsecS:nullFI}

Substituting Eq.~(\ref{eqS:nulllogderivative}) into Eq.~(\ref{eqS:FITRY}) gives
\begin{equation}
F_{\theta}^{\mathrm{TRY}}(x_0)=\Var_{Y|x_0}
\left[\frac{2\,\mathrm{Re}\!\left[\epsilon^*(y)q(y)\right]}{|\epsilon(y)|^2+b}
\right]_{\theta=0}.\label{eqS:nullFI}
\end{equation}
The background floor $b$, finite aperture, finite source discretization, and finite photon number all keep Eq.~(\ref{eqS:nullFI}) finite. This is why the main text uses the conservative language ``enhanced Fisher information per detected event'' rather than ``divergent'' or ``unbounded'' Fisher information.

\begin{figure}[t]
\includegraphics[width=\columnwidth]{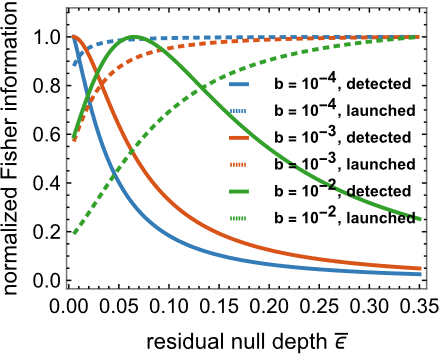}
\caption{
Regularized null-response Fisher information for different background floors $b$. Solid curves show the conditional Fisher information per detected event, normalized by the maximum value for each $b$. Dashed curves show the conservative per-launched-photon contribution obtained by multiplying the conditional Fisher information by the detection probability. The plot illustrates the central tradeoff: operating near a dark response can enhance conditional sensitivity, but the total photon-budget advantage is limited by background and event rate.
}
\label{figS:nullreg}
\end{figure}

Figure~\ref{figS:nullreg} shows this regularization explicitly. For each background floor $b$, the solid curve is the conditional Fisher information per detected event, normalized by its own maximum. The dashed curve is the conservative per-launched-photon contribution $P_{\mathrm{det}}F_{\theta}^{\mathrm{cond}}$, also normalized by its own maximum. As the residual null depth is reduced, the conditional Fisher information can increase, but the detected event rate decreases. The figure therefore supports the main-text statement that null operation concentrates information into the conditional record without creating an unlimited photon-budget advantage.

\section{Partially coherent and thermal fields}
\label{secS:coherence}

\subsection{Coherence matrix and coherence entropy}
\label{subsecS:cohmatrix}

For a partially coherent scalar field in the source plane, define the cross-spectral density or coherence matrix
\begin{equation}
\Gamma(y,y')=\left\langle E^*(y)E(y')
\right\rangle,\label{eqS:Gamma}
\end{equation}
where the brackets denote an ensemble average. The normalized coherence operator is
\begin{equation}
\rho_{\coh}=\frac{\Gamma}{\Tr(\Gamma)}.
\label{eqS:rhocoh}
\end{equation}
Its coherence entropy is
\begin{equation}
S_{\coh}=-\Tr\!\left(\rho_{\coh} \ln\rho_{\coh}\right)=-\sum_n \lambda_n\ln\lambda_n,\label{eqS:Scoh}
\end{equation}
where $\lambda_n$ are the eigenvalues of $\rho_{\coh}$. This is the optical analogue of the von Neumann entropy of the normalized coherence operator.

Under ideal lossless propagation,
\begin{equation}
\rho_{\coh}\rightarrow U\rho_{\coh}U^{\dagger},
\label{eqS:unitary}
\end{equation}
where $U$ is the propagation operator. The eigenvalues $\lambda_n$ are unchanged by this unitary similarity transformation, and therefore $S_{\coh}$ is unchanged. This is the precise technical reason the main text avoids saying that TRY reverses thermal entropy.

\subsection{Filtering, postselection, and entropy change}
\label{subsecS:filter}

Aperture truncation, loss, conditioning, and detector postselection are nonunitary operations. A generic filter $M$ maps
\begin{equation}
\rho_{\coh}\rightarrow\rho_{\coh}'=
\frac{M\rho_{\coh}M^{\dagger}}
{\Tr(M\rho_{\coh}M^{\dagger})}.\label{eqS:filter}
\end{equation}
The eigenvalues of $\rho_{\coh}'$ need not equal those of $\rho_{\coh}$, so the coherence entropy can change. Therefore, entropy changes in a realistic TRY experiment should be attributed to filtering, coarse graining, conditioning, loss, or postselection, not to a reversal of thermodynamic time. This supports the main-text statement that TRY reorganizes measured conditional entropy rather than reversing thermodynamic entropy.

\begin{figure}[t]
\includegraphics[width=\columnwidth]{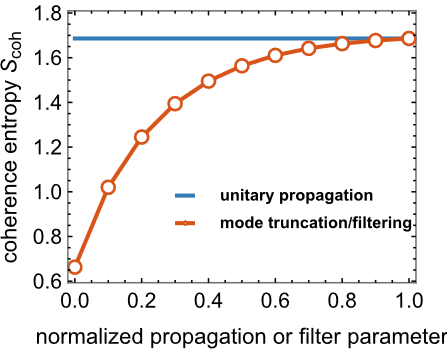}
\caption{
Coherence entropy under two idealized operations. A unitary propagation preserves the eigenvalues of the normalized coherence matrix and therefore leaves $S_{\mathrm{coh}}$ unchanged. Mode truncation or filtering changes the eigenvalue spectrum after renormalization and can change the coherence entropy. This illustrates why entropy changes in TRY should be associated with conditioning, filtering, or coarse graining rather than with a reversal of thermodynamic entropy.
}
\label{figS:coherence}
\end{figure}

Figure~\ref{figS:coherence} illustrates this distinction. The unitary curve remains flat because the modal eigenvalues are preserved. The filtering curve changes because truncation followed by renormalization changes the modal spectrum. The figure is not intended to model a specific thermal source; it is a diagnostic example showing why entropy changes require a nonunitary operation or a change in the observed degrees of freedom.

\section{Numerical recipes for the figures}
\label{secS:numerics}

The figures in the main text and this supplement are illustrative dimensionless calculations designed to visualize the statistical structure of the theory. They are not intended to fit a particular experimental apparatus. The purpose of the numerical examples is to show how the same reciprocal kernel generates different marginal, conditional, and information-theoretic quantities.

\subsection{Reciprocal two-path kernel}
\label{subsecS:numkernel}

For Fig.~2 of the main text and Figs.~\ref{figS:kernel}--\ref{figS:KL}, we used the dimensionless two-path amplitude
\begin{align}
&K_{\beta}(x,y)=\exp\!\left[-\frac{(x-y-d/2)^2}{2\sigma^2}+i\beta (x-y-d/2)^2\right]\nonumber\\
&\quad+\exp\!\left[-\frac{(x-y+d/2)^2}{2\sigma^2}
+i\beta (x-y+d/2)^2\right],\label{eqS:numericalK}
\end{align}
with
\begin{equation}
d=1.3,\quad\sigma=0.65,\quad x,y\in[-3,3].
\label{eqS:numparameters}
\end{equation}
The fixed detector for the TRY conditional distribution is $x_0=0$. The dimensionless parameter $\beta$ is the coefficient of the quadratic phase terms in Eq.~(\ref{eqS:numericalK}). It may be viewed as a defocus-like realization of the more general perturbation parameter $\theta$ in the main text, but it is introduced here only to visualize how the different information measures respond to a controlled tuning of the reciprocal kernel.

The joint distribution was defined on a uniform square grid by
\begin{equation}
p_{\beta}(x,y)=\frac{|K_{\beta}(x,y)|^2}
{\sum_{x,y}|K_{\beta}(x,y)|^2}.\label{eqS:numP}
\end{equation}
The standard marginal, TRY conditional distribution, detector entropy, source-label entropy, and mutual information were then computed from their discrete definitions. In the main-text Fig.~2, $\beta$ was varied over $0\leq\beta\leq2.5$. The figure should be read as a structural comparison: $H_X$ is a detector marginal entropy, $H_{Y|x_0}$ is a fixed-detector source conditional entropy, and $I(X;Y)$ is the source--detector mutual information computed from the full joint distribution.

\subsection{Null-response calculation}
\label{subsecS:numnull}

For Fig.~3 of the main text and Fig.~\ref{figS:nullreg}, we used the real-valued regularized model
\begin{equation}
R_{\theta}(y)=\left[\bar{\epsilon}+0.12y^2+\theta q(y)\right]^2+b,\label{eqS:numericalNull}
\end{equation}
where
\begin{equation}
q(y)=\sin(\pi y)+0.35y.\label{eqS:numericalq}
\end{equation}
The source coordinate was sampled uniformly on $-1\leq y\leq1$. For each value of the residual null-depth parameter $\bar{\epsilon}$, the conditional distribution was
\begin{equation}
p(y|\mathrm{det})=\frac{R_0(y)}{\sum_yR_0(y)}.
\label{eqS:numericalCond}
\end{equation}
The conditional Fisher information was evaluated from the discrete version of Eq.~(\ref{eqS:FITRY}). The detection probability was represented by the grid average of $R_0(y)$. In the main-text Fig.~3, the background floor was $b=10^{-3}$ and $\bar{\epsilon}$ was varied from $0.005$ to $0.35$. Figure~\ref{figS:nullreg} repeats the calculation for $b=10^{-4}$, $10^{-3}$, and $10^{-2}$ to show how background regularizes the conditional Fisher information and the per-launched-photon contribution.

\subsection{Coherence-entropy calculation}
\label{subsecS:numcoh}

For Fig.~\ref{figS:coherence}, we used a normalized modal spectrum
\begin{equation}
\lambda_n=\frac{e^{-n/2}}{\sum_{m=0}^{N-1} e^{-m/2}},\quad n=0,\ldots,N-1,\label{eqS:lambdas}
\end{equation}
with $N=12$. A unitary transformation preserves this spectrum and hence preserves $S_{\coh}$. A simple truncation model retains only the first $M$ modes, renormalizes the remaining eigenvalues, and recomputes Eq.~(\ref{eqS:Scoh}). This gives the filtered curve in Fig.~\ref{figS:coherence}. The purpose is to illustrate the difference between entropy-preserving unitary propagation and entropy-changing filtering or postselection.

\section{Summary of what is established in the supplement}
\label{secS:summary}

The technical results above support the main-text claims as follows. First, Eqs.~(\ref{eqS:generalK})--(\ref{eqS:tryconditional}) show that standard Young and TRY are different statistical reductions of the same reciprocal source--detector kernel. Second, Eqs.~(\ref{eqS:MI})--(\ref{eqS:KLidentity}) show that marginal entropy is not the reciprocal invariant; mutual information is. Third, Eqs.~(\ref{eqS:score})--(\ref{eqS:FITRY}) show that the TRY Fisher information is the source-space variance of a logarithmic fixed-detector response. Fourth, Eqs.~(\ref{eqS:nullresponse})--(\ref{eqS:nullFI}) show how null-response sensitivity is regularized by background and finite event rate. Finally, Eqs.~(\ref{eqS:Gamma})--(\ref{eqS:filter}) clarify why the manuscript concerns measured conditional entropy and coherence entropy, not a reversal of thermodynamic entropy.